\documentclass[aps,prb,twocolumn,floatfix]{revtex4}
\usepackage{bm}
\usepackage{epsf}
\usepackage{amssymb}
\usepackage{amsmath}
\usepackage{graphicx}
\usepackage{rotating}
\usepackage{epsfig}
\usepackage{psfrag}
\usepackage{amsmath}
\usepackage{hyperref}
\usepackage{subfigure}
\usepackage{braket}
\usepackage{tikz}
\usepackage{stmaryrd}
\usepackage{wasysym}
\usepackage{xcolor}

\begin{document}
\title{Revival of antibiskyrmionic magnetic phases in bilayer NiI$_2$}

\author{Jyotirish Das$^1$, Muhammad Akram$^{1,2}$, Onur Erten$^1$}
\affiliation{$^1$Department of Physics, Arizona State University, Tempe, AZ 85287, USA \\ $^2$Department of Physics, Balochistan University of Information Technology, Engineering and Management Sciences (BUITEMS), Quetta 87300, Pakistan}

\begin{abstract}
{Magnetic skyrmions are topologically protected spin textures with potential applications in memory and logic devices. Skyrmions have been commonly
observed in systems with Dzyaloshinskii–Moriya interaction due to broken inversion symmetry. 
Yet, recent studies suggest that skyrmions can also be stabilized in systems with inversion symmetry such as Ni-based dihalides due to magnetic frustration. In this article, we employ atomistic simulations to investigate chiral magnetic phases in bilayers of NiI$_2$ and NiBr$_2$. We show that the antiferromagnetic interlayer coupling introduces an additional magnetic frustration and gives rise to a variety of novel spin textures with different topological charges. Specifically for NiI$_2$, we observe that the skyrmions with the in-plane component of spins wrapping around twice (biskyrmions) have an enhanced stability compared to the monolayer case. We also study the polarization induced by the non-colinear magnetic order in NiI$_2$ bilayers and show that the polarization of the topologically nontrivial phases is negligible compared to the spiral phases. Thus, we conclude that polarization measurements can be an indirect route for detecting skyrmions in upcoming experiments.
}
\end{abstract}
\maketitle

\section{Introduction}
Nickel dihalides, NiX$_2$ (X = I, Cl, Br), belong to a class of insulating van der Waals (vdW) magnets with transition temperatures ranging from $T_c = 52$K in NiCl$_2$ and NiBr$_2$ up to $T_c = 76$K in NiI$_2$ in the bulk\cite{doi:10.1021/acsomega.9b00056,PhysRevMaterials.3.044001, cryst7050121}. They exhibit a variety of magnetic phases including ferromagnetic (FM), antiferromagnetic (AFM) and spiral (Sp) ground states\cite{cryst7050121}. The Sp phases in NiI$_2$ and NiBr$_2$ are of particular interest as they break inversion symmetry and thus exhibit finite polarization, leading to multiferroic properties\cite{cryst7050121, PhysRevB.84.060406}. Monolayers of NiX$_2$ can be obtained by suitable exfoliation methods and both magnetism and multiferroic properties survive down to single layer\cite{Song_Nature2022, doi:10.1021/acs.nanolett.1c01095}. Recent theoretical studies\cite{amoroso2020spontaneous} predict that NiX$_2$ monolayers may also host chiral magnetic phases such as skyrmions (SkX) and antibiskyrmions (A2Sk) even in the absence of Dzyaloshinskii-Moriya interaction (DMI). Skyrmions are topologically protected vortex-like magnetic textures with potential applications in logic and memory devices\cite{doi:10.1021/acs.chemrev.0c00297, fert,ever,bogdanov}.
In nickel dihalides, these phases are predicted to be stabilized by a combination of magnetic frustration, anisotropic exchange\cite{amoroso2020spontaneous} and external magnetic field (B). However, atomistic simulations for NiI$_2$ monolayer indicate that the A2Sk and Sp phases are quite close in energy\cite{PhysRevMaterials.7.054006}. Indeed, circular dichroic Raman measurements show that multiferroic order persists at monolayers, consistent with a Sp ground state\cite{Song_Nature2022}. Biskyrmions have twice the topological charge of skyrmions. Experimentally, they have only been observed in a handful of compounds including La$_{2-2x}$Sr$_{1+2x}$Mn$_2$O$_7$ \cite{Yu2014}, MnNiGa \cite{https://doi.org/10.1002/adma.201600889}, Cr$_{11}$Ge$_{19}$ \cite{PhysRevLett.120.037203}, MnPdGa\cite{10.1063/1.5089609} and Nd$_2$Co$_{17}$ \cite{Zuo2023}. However some of these observations can be misleading since the topologically trivial magnetic bubbles can show similar images under Lorentz microscopy\cite{Loudon, Yao2019, PhysRevB.103.214435}. In light of these observations, it is important to find new platforms which can stabilize biskyrmions.
\begin{figure}[h!]  \includegraphics[width=0.5\textwidth]{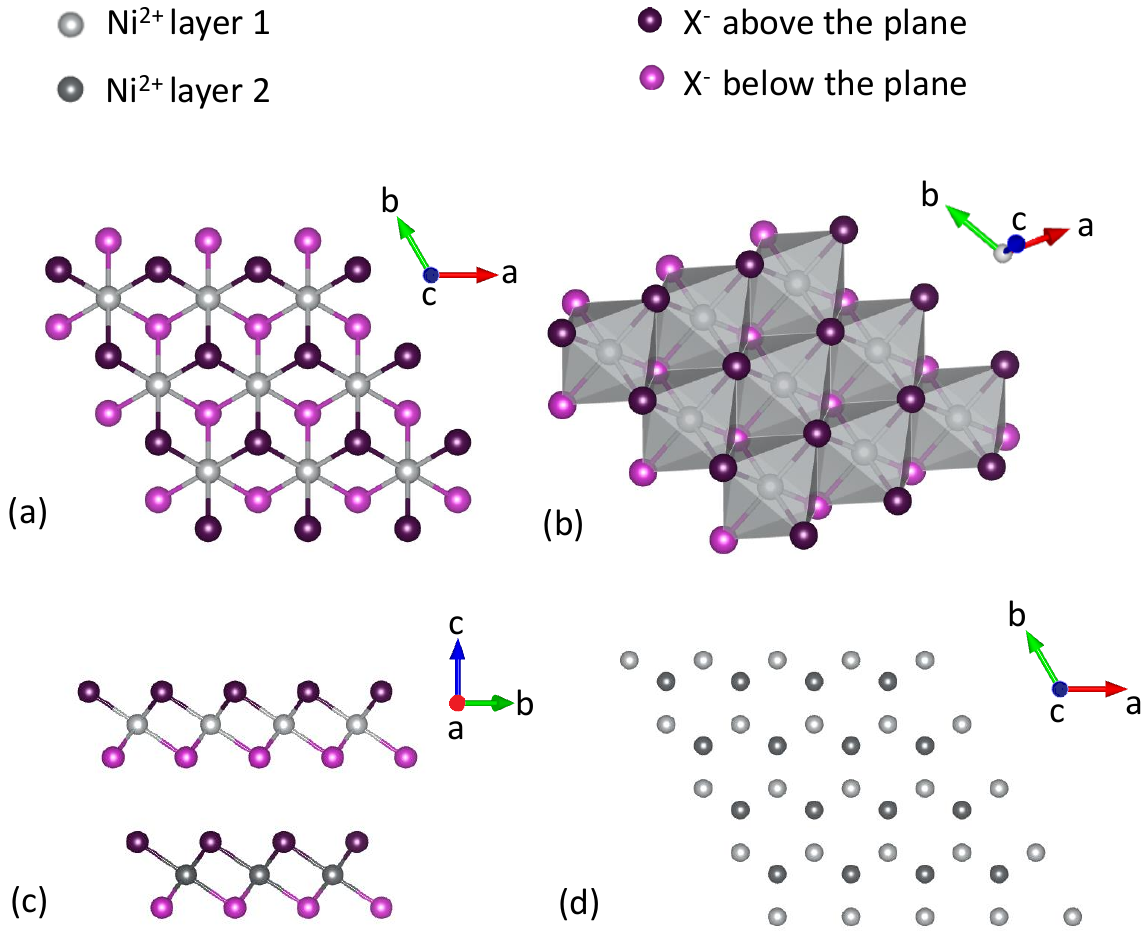}
    \caption{(a) Front view and (b) side view of NiX$_2$ (X = I, Br) monolayer. NiX$_2$ bilayer with rhombohedral stacking, (c) side view with X atoms and (d) front view showing only the halide atoms.}
    \label{atoms}
\end{figure} 

Due to weak interlayer bonding, vdW materials can be arranged in different stacking patterns and further manipulated through twisting to create moir\'e superlattices\cite{Bistritzer_PNAS2011, Cao2018}. In both cases, completely new phenomena that is not possible to obtain in monolayers can be achieved\cite{He_ACS2021}. A range of new non-coplanar phases are predicted\cite{Tong_ACS2018, Hejazi_PNAS2020, Hejazi_PRB2021, Akram_PRB2021, Akram_NanoLett2021, Tong_PRR2021} for moir\'e magnets and some of these phases have been observed experimentally\cite{Xu_NatNano2021, Song_Science2021, xie2022}. For NiX$_2$ bilayers, interlayer interactions are antiferromagnetic and therefore compete with the external magnetic field and introduce additional magnetic frustration. Motivated by this observation, we study the phase diagram of NiI$_2$ and NiBr$_2$ bilayers in rhombohedral and AA stacking patterns via atomistic Landau-Lifshitz-Gilbert (LLG) simulations as a function of interlayer exchange and external magnetic field. Our main results show that: 
(i) for NiI$_2$, the interlayer coupling narrows the region of the SkX phase and promotes antibiskyrmionic phases with different topological charges for both AA and rhombohedral stacking orders. (ii) As a result, for the range of {\it ab initio} estimates for interlayer exchange\cite{doi:10.1021/acsnano.0c04499, Song_Nature2022}, it is feasible to stabilize A2Sk phases in NiI$_2$ bilayers compared to monolayer systems. (iii) Larger values of interlayer exchange leads to the complete suppression of A2Sk/SkX phases and gives rise to the formation of Sp phases instead. (iv) Due to negligible anisotropic exchange , skyrmions are completely suppressed in NiBr$_2$ for realistic interlayer exchange parameters. (v) In NiI$_2$, the electric polarization induced by the non-colinear magnetic order increases with the magnetic field in the Sp phase and is negligible in the A2Sk/SkX phases. This prediction can be applied to deduce the skyrmionic phases indirectly.

The rest of the article is organized as follows. First, we introduce the effective spin Hamiltonian and overview the results of atomistic simulations for the NiI$_2$ monolayer. Next, we discuss the phase diagram of NiI$_2$ and NiBr$_2$ bilayers in rhombohedral and AA stacking patterns. We conclude with a discussion on the induced polarization due to magnetic order and experimental signatures of the phase diagram.

\section{Microscopic magnetic model}
NiX$_2$ belongs to $R \bar{3} m$ space group which is a centro-symmetric rhombohedral structure. In a monolayer, Ni$^{2+}$ ions (3d$^8$, S=1) form a triangular lattice with X$^-$ (X=I, Cl, Br) ions are arranged above and below the plane as shown in Fig. \ref{atoms}. The magnetic interactions between localized Ni spins ($\mathbf{s}_{i}^{l}$) bilayers can be modeled by the following spin Hamiltonian\cite{amoroso2020spontaneous}: 
\begin{eqnarray}
    H&=& \frac{1}{2} \sum_{i \neq j,l} \mathbf{s}_{i}^{l} \cdot \mathbf{J}_{ij} \cdot \mathbf{s}_{j}^{l} +J^{\perp} \sum_{\langle i j \rangle} \mathbf{s}_{i}^{1} \cdot \mathbf{s}_{j}^{2} \nonumber \\ &+& 
    \sum_{i,l} \mathbf{s}_{i}^{l}  \cdot \mathbf{A}_{i} \cdot \mathbf{s}_{i}^{l}  - \sum_{i,l}\mathbf{B} \cdot  \mathbf{s}_{i}^{l}   \, .
    \label{eq:1}
\end{eqnarray}
Here, $l = 1,2$ denotes the layer index. \textbf{J}$_{ij}$ is the tensor for the exchange coupling interactions which can be  decomposed into an isotropic coupling term and an anisotropic one, the latter also referred to as the two-site anisotropy. $J^{\perp}$ is the antiferromagnetic interlayer exchange, \textbf{A}$_{i}$ is the single-ion anisotropy and $\mathbf{B}$ is the external magnetic field. For the intralayer exchange parameters, we use the values obtained from first principle calculations in Ref. \citenum{amoroso2020spontaneous} (See Appendix A for more details).

The magnetic frustration in nickel dihalides originate from a strong ferromagnetic nearest-neighbor exchange interaction
(J$^{\rm 1iso}$) with a comparable antiferromagnetic third nearest-neighbor exchange (J$^{\rm 3iso}$). Theoretical studies show that the interplay of the magnetic frustration and exchange anisotropy is the key ingredient for the stabilization of topologically protected spin textures in monolayer NiX$_2$\cite{amoroso2020spontaneous, PhysRevMaterials.7.054006}.

\section{Results and Discussion}

\begin{figure}
    \includegraphics[width=0.45\textwidth]{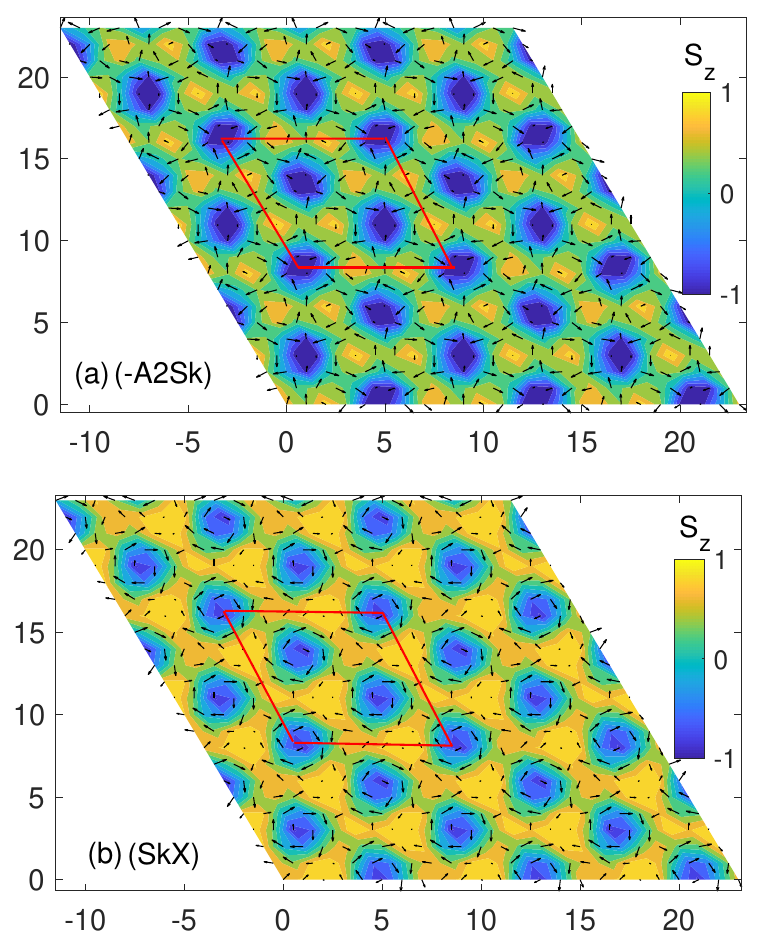}
    \caption{Skyrmion and antibiskyrmion phases in NiI$_2$ monolayer: (a)-A2Sk at B$_z$ = 36.2 T and (b) SkX at B$_z$ = 18.1 T. The area marked by red denotes the magnetic unit cell. The arrows denote the in-plane component while the colormap denotes the out-of-plane component of the spin.}
    \label{sk_35}
\end{figure}

\subsection{Overview of the phase diagram of NiI$_2$ monolayer}
The phase diagram of monolayer NiI$_2$ (eq.~\ref{eq:1}) has been studied via Monte Carlo simulations\cite{amoroso2020spontaneous}. In a previous work\cite{PhysRevMaterials.7.054006}, we have extended these results to the Janus counterparts via atomistic LLG simulations. The magnetic unit cell ($L\times L$) for these simulations is estimated by the Luttinger-Tisza (LT) method \cite{Luttinger_PR1946} and we verify these values by performing system-size dependent calculations (see Appendix C for more details).
Our atomistic simulations show that magnetic field up to 29.3T give rise to a Sp phase, which is energetically very close to an -A2Sk phase, ($\Delta E < 0.05~ \rm meV$). For intermediate magnetic fields, $29.3 T < B < 69.9T$, we obtain a SkX phase. A further increase in the magnetic field results in a Sp phase which adiabatically connects to a ferromagnet. 

The different magnetic phases are distinguished by the topological charge Q measured over an $8 \times 8$ magnetic unit cell. A2Sk, SkX and Sp phases have $Q = 6$, $3$, $0$ respectively. Representative SkX and A2Sk spin textures are shown in Fig. \ref{sk_35}. The spin structure factor $S({\bf q})$ is also a useful indicator to distinguish different phases. 

\subsection{Magnetic phase diagram of NiI$_2$ bilayer}

\begin{figure}
    \includegraphics[width=8.4 cm]{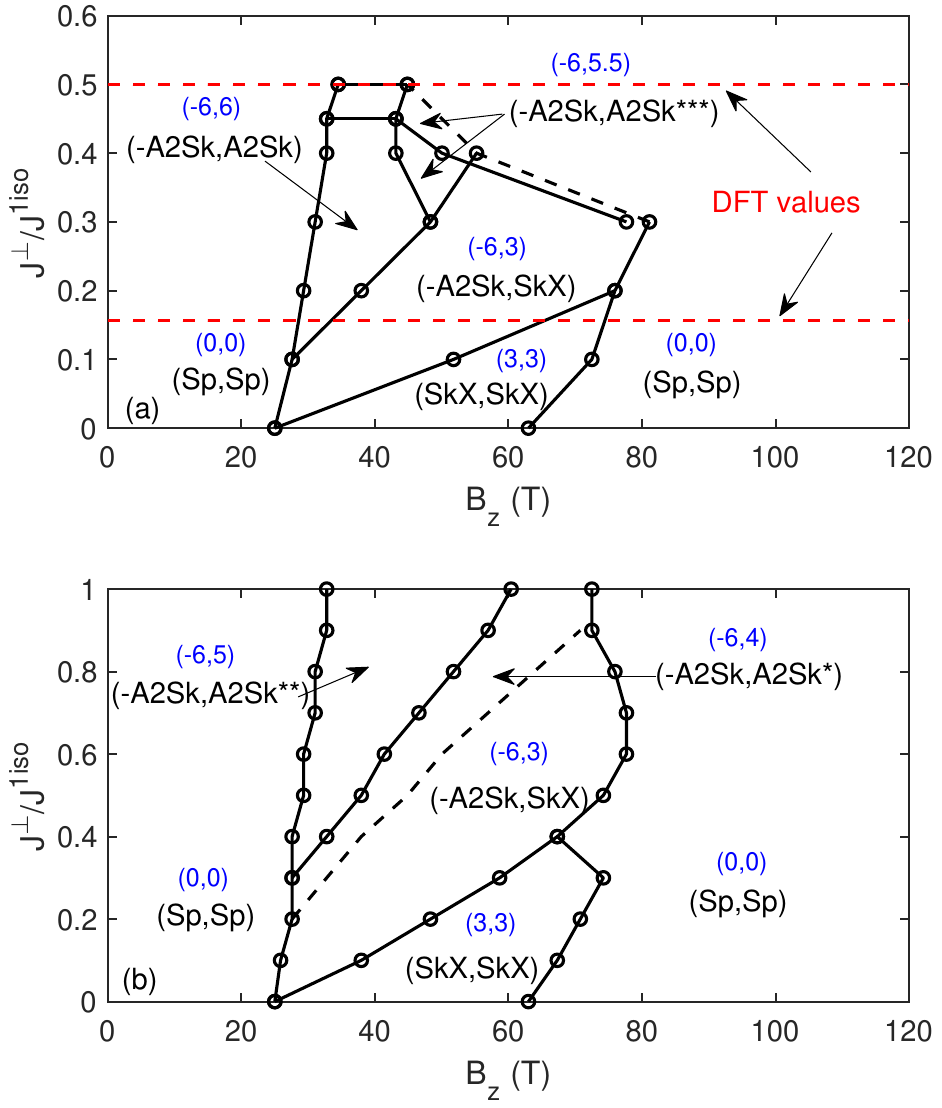}
    \caption{The phase diagram of NiI$_2$ bilayer for (a) rhombohedral and (b) AA stacking. The dashed red lines indicate the {\it ab initio} values for the interlayer exchange obtained from Refs. \citenum{doi:10.1021/acsnano.0c04499} and \citenum{Song_Nature2022}. The black dashed lines imply the regions of quasi-degenerate solutions with energy difference less than $0.01$ meV. The SkX and A2Sk phases are sandwiched between the Sp phases. Interlayer exchange suppresses the SkX phase leading to a revival of antibiskyrmionic phases.}
    \label{NiI2_ab}
\end{figure}

Even though NiI$_2$ has rhombohedral stacking in the bulk, {\it ab initio} calculations show that the energy differences between the rhombohedral and AA stacking patterns for bilayer systems are relatively small \cite{Wang2022}, $\Delta E \simeq 0.3$ meV. Therefore, we study the magnetic phase diagrams of both stacking orders by using atomistic LLG simulations, following a method similar to the monolayer case. The LT method shows that the antiferromagnetic interlayer exchange does not affect the size of the magnetic unit cell. The phase diagram of NiI$_2$ for rhombohedral and AA stacking patterns are presented in Fig. \ref{NiI2_ab}(a) and (b) respectively. The dashed red lines in Fig. \ref{NiI2_ab}(a) indicate the two values for interlayer coupling obtained from {\it ab initio} calculations in Refs.  \citenum{doi:10.1021/acsnano.0c04499} and \citenum{Song_Nature2022}. To the best of our knowledge, there is no {\it ab initio} estimate for the interlayer exchange in AA stacking. The magnetic ground states and the corresponding topological charge on each layer are indicated. 
The phase -A2Sk is an antibiskyrmion with a topological charge $Q=-6$ per unit cell. A2Sk$^*$, A2Sk$^{**}$ and A2Sk$^{***}$ are antibiskyrmionic intermediate phases with varying topological charge $4,~5$ and $5.5$. We call these phases antibiskyrmionic since the spin texture closely resembles the A2Sk phase.
For both stacking patterns, the phase diagrams show that topologically nontrivial phases such as SkX and A2Sk are sandwiched between the Sp phases. There are two distinguishing features of this phase diagram. Firstly, the antiferromagnetic interlayer coupling enables the revival of the A2Sk phase which was suppressed by the Sp phase in the monolayer case. This comes at the cost of inhibiting the SkX phase at an interlayer coupling strength of $J^\perp/J^{1\rm{iso}}\sim$ 0.2 and 0.4 for the rhombohedral and AA stackings, respectively. Secondly, at larger values, the interlayer coupling completely suppresses the antibiskyrmionic phases in favor of the Sp phase. Next, we elucidate these findings and other features of the phase diagram.
\begin{table}[!t]
\begin{center}
\begin{tabular}{||c | c |c |c |c |c |c  | c ||} 
 \hline
  & $E$ & $E^1$ & $E^3$ & $E^{2an}$ & $E^b$& $E^{I}_{\parallel}$ & $E^{I}_{\perp}$\\ [0.5ex] 
 \hline\hline
 (-A2Sk, SkX) & -16.92 & -10.35 & -3.36 & -1.22 & -1.63 & -0.48 & -0.09 \\ 
 \hline
 (SkX, SkX) & -16.89 & -10.57 & -3.04 & -1.21 & -1.8 & -0.37 & -0.09 \\  
 \hline
 $\Delta E$ & -0.03 & 0.22 & -0.32 & -0.01 & 0.17 & -0.11 & 0.00 \\ [1ex] 
 \hline
\end{tabular}
\caption{\label{e_table} Energy contributions from different terms of the spin Hamiltonian for rhombohedral stacking at 46.6 T and J$^{\perp}$/J$^{1\rm{iso}} = 0.1 $ in units of meV. $E$ is the total energy per site. $E^1$ ($E^3$) is the energy due to (anti-) ferromagnetic exchange, $E^{2an}$ represents the energy due to the two-site anisotropy, $E^{I}_{\parallel}$ ($E^{I}_{\perp}$) is the energy due to the parallel (perpendicular) component of the interlayer exchange, $E^b$ is the energy contribution due to an external magnetic field. Energies due to single-site anisotropy and due to interaction with the second nearest neighbour have been ignored. $\Delta E$ is defined as $\Delta E = E_{(-A2Sk,SkX)}-E_{(SkX,SkX)}$.}
\end{center}
\end{table}

\subsubsection{Revival of the antibiskyrmionic phases}
A key difference between the monolayer and bilayer phase diagrams of NiI$_2$ is the restoration of the antibiskyrmionic phases at intermediate $B$. Fig. \ref{NiI2_ab}(a) and (b) show that this effect can be observed in both stacking orders and it is primarily due to the suppression of the (SkX, SkX) phase with interlayer coupling. In order to illustrate the competition between (SkX, SkX) and (-A2Sk, SkX) phases and the effects of interlayer exchange and B in detail, we tabulate the total energy of the bilayer and its various contributions from the spin Hamiltonian for both the competing phases for B$_z$ = 46.6 T and J$^{\perp}$/J$^{1\rm{iso}} = 0.1 $ in Table 1. For these parameters, the ground state is (-A2Sk, SkX), and we recover (SkX, SkX) as a local minimum in LLG simulations. We find that the intralayer exchange interaction $E^1$ ($E^3$) stabilizes the SkX (-A2Sk) phase as it has effectively a greater number of nearest neighbor spins which are aligned (anti-)parallel to each other. Moreover the contribution to the energy due to the external magnetic field ($E^b$) stabilizes the (SkX, SkX) phase as it has a larger magnetization along the $\hat{z}$ direction compared to (-A2Sk, SkX). Yet the in-plane component of interlayer interaction ($E^{I}_{\parallel}$) stabilizes the (-A2Sk, SkX) phase. As a result, for large interlayer exchange antibiskyrmionic phases takes over the (SkX, SkX) phase. 

\begin{figure}[t!]
  \centering
  \includegraphics[width=0.4\textwidth]{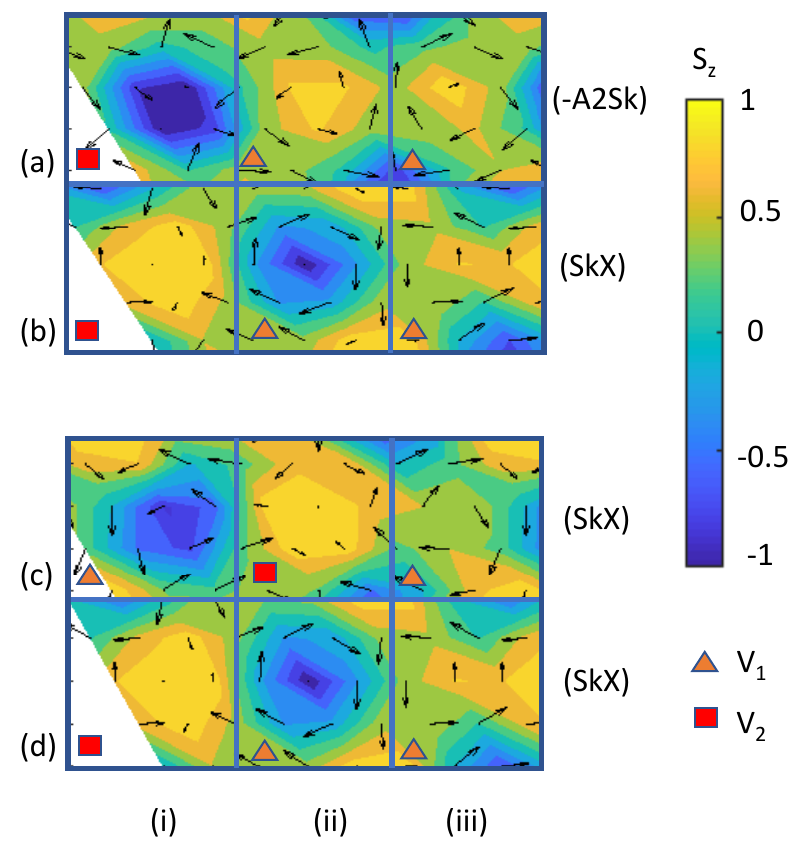}
  \caption{Spin textures of bilayer NiI$_2$ in rhombohedral stacking for $B_z= 46.6$ T and $J^{\perp}=0.1 J^{1iso}$. (a) and (b) denotes the (-A2Sk, SkX) phase whereas (c) and (d) denotes the (SkX, SkX) phase respectively.}
  \label{cross}
\end{figure}

$E^{I}_{\parallel}$ stabilizes the (-A2Sk, SkX) phase due to the relative arrangement of the skyrmions and antibiskyrmions within each layer. This can be understood more intuitively with the help of the spin textures for the two layers in both the competing phases as shown in Fig.~\ref{cross}. Spin textures consist of a periodic pattern of vortices which have skyrmionic/antibiskyrmionic structures but with fractional topological charges\cite{amoroso2020spontaneous}. We call the vortices which wind in the plane once (hence with a skyrmionic structure) V$_1$ and the ones which wind in the opposite direction twice with an antibiskyrmionic structure V$_2$ for convenience. Nine such vortices form the smallest repetitive spin texture within the magnetic unit cell (see Appendix B). 
In Fig. \ref{cross}, we compare the six vortices in the (-A2Sk, SkX) phase ((a), (b)) with those in the (SkX, SkX) phase ((c),(d)) at a magnetic field of 46.6 T and $J^{\perp}/J^{1\rm iso}=0.1$. In the (-A2Sk, SkX) phase ((a), (b)) vortices of the same type are arranged on top of each other while this is not the case for the (SkX, SkX) phase ((c), (d)). This is because in the (SkX, SkX) phase, vortices of type V$_2$ have the out-of plane component of the magnetization pointing in the same direction i.e both of them have the same polarity. Therefore, in this phase the antiferromagnetic interlayer interaction couples the vortex V$_2$ to a vortex V$_1$ with opposite polarity. However, the coupling of similar vortices in the (-A2Sk, SkX) phase result in a better alignment of the in-plane spin components thereby lowering their in-plane interlayer coupling energy ($E_{\parallel}^{I}$) compared to the (SkX, SkX) phase. 

\begin{figure}[t!]
    \includegraphics[width=7.0cm]{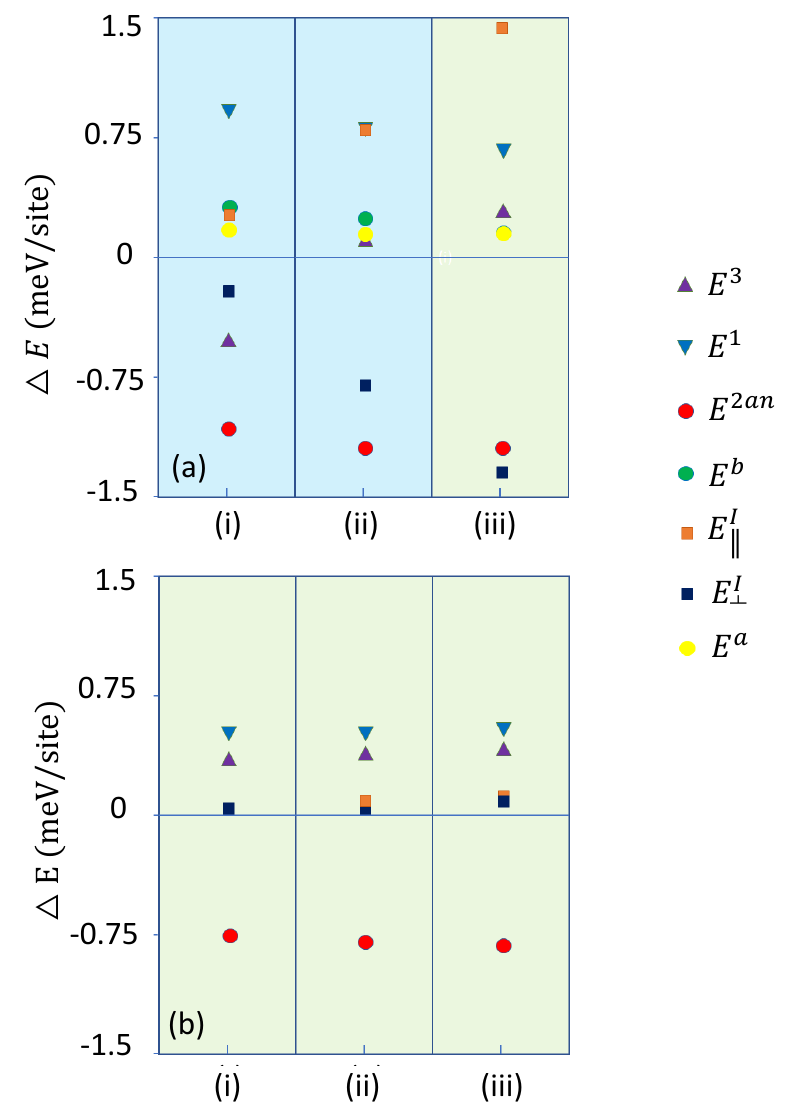}
    \caption{represents the energy difference $\Delta E= E_{(-A2Sk, {\rm X})}-E_{(Sp, Sp)}$ between different skyrmionic phases (-A2Sk, X) and the spiral phase for rhombohedral stacked NiI$_2$. The different parameters are for (a) B$_z$ = 46.6 T, (i) X= SkX, J$^{\perp} /J^{1\rm iso}$ = 0.1, (ii) X= A2Sk***, J$^{\perp} /J^{1\rm iso}$ = 0.4, (iii) X= A2Sk, J$^{\perp} /J^{1\rm iso}$ = 0.7 and for (b) B$_z$ = 13.8 T and X= A2Sk at J$^{\perp} /J^{1\rm iso}$ = 0.1 , 0.4 and 0.7 for (i), (ii) and (iii) respectively. Blue background indicates (-A2Sk, X) ground state whereas green background indicates (Sp, Sp) ground state.  
    $E^1$ ($E^3$) is the energy due to (anti-) ferromagnetic exchange, $E^{2an}$ represents the energy due to the two-site anisotropy, $E^{I}_{\parallel}$ ($E^{I}_{\perp}$) is the energy due to the parallel (perpendicular) component of the interlayer exchange, $E^b$ is the energy contribution due to an external magnetic field and $E^a$ is the energy contribution due to single-site anisotropy.}
    \label{ed1}
\end{figure}

\subsubsection{Suppression of the antibiskyrmionic phase}
In order to illustrate the mechanism behind the suppression of antibiskyrmionic phases for favor of Sp phase at large $J^\perp$, we present the energy difference between these phases for various terms of the spin Hamiltonian for rhombohedral stacking in Fig.~\ref{ed1}. We observe that keeping the magnetic field constant at B$_z$ = 46.6 T, an increase in interlayer coupling from J$^{\perp}$/J$^{1\rm iso}$ = 0.1 (Fig. \ref{ed1}(a)(i)) to J$^{\perp}$/J$^{1\rm iso}$ = 0.7 (Fig. \ref{ed1}(a)(iii)) changes the ground state from skyrmions (blue background) to spirals (green background). The dominating contributions result from the energy difference due to two-site anisotropy ($E^{2an}$) and the parallel ($E^{I}_{\parallel}$) and perpendicular ($E^{I}_{\perp}$) components of interlayer coupling energy. We observe that $\Delta E^{I}_{\parallel}$ is slightly greater in magnitude than $\Delta E^{I}_{\perp}$ and both of them increases almost linearly with interlayer coupling strength. $\Delta E^{2an}$ by contrast tends to saturate and cannot compensate for $\Delta E^{I}_{\parallel}$ which destabilizes the chiral phases at high values of interlayer coupling. We also observe that the energy due to antiferromagnetic exchange ($E^3$) stabilizes skyrmions for lower values of interlayer coupling but stabilizes spirals at higher values. We therefore conclude that the parallel component of interlayer coupling energy and the energy due to antiferromagnetic exchange are the dominant factors which tends to destabilize chiral phases and form spirals at high values of interlayer coupling.

\subsubsection{Other features of the phase diagram}
Fig.\ref{ed1}(b)(i)-(iii) shows the energy decomposition for rhombohedral stacking for J$^{\perp}/J^{1\rm iso}$ = 0.1, 0.4 and 0.7  respectively at  B$_z$ = 13.8 T. We observe that the different energy contributions show negligible variation with varying interlayer coupling at relatively small magnetic fields. This even includes the energy contributions due to the interlayer coupling. Therefore, the phase boundary between the spiral phases and the antibiskyrmionic phases is nearly vertical for both rhombohedral and AA stacking at $B \sim$ 22 T. On the contrary, larger values of interlayer exchange forces the spins on each layer to be oriented anti-parallel. This increases the polarity and thereby increasing the topological charge. Therefore, we observe that the (-A2Sk, SkX) phase undergoes a transition into the (-A2Sk, A2Sk) phase for rhombohedral stacking and into the (-A2Sk, SkX*) and (-A2Sk, SkX**) phases successively for AA stacking at $B\sim$ 34.5 T. The opposite phenomenon is observed when the magnetic field is increased keeping the interlayer coupling constant. A strong magnetic field acts against the interlayer coupling and forces the spins of both the layers to be oriented along magnetic field which facilitates successive transitions into phases with lower topological charges. 

It is important to note that for rhombohedral stacking, we obtain a phase (-A2Sk, A2Sk$^{***}$) which has a fractional topological charge, $Q=5.5$, on one layer (A2Sk$^{***}$). This is due to the fact that the topological charge is calculated for each layer separately. Therefore it does not consider the bilayer system as a whole and ignores the chirality arising from the interlayer coupling of the spins. When we consider the bilayer system as effectively one layer with two antiferromagnetically coupled sublattices, we get some additional triangular plaquettes. If we reverse the spins of one layer to take the antiferromagnetic coupling into account and sum over these additional triangular plaquettes (see Fig.~\ref{lat2}),  a net topological charge $Q = 6$ for a magnetic unit cell of the entire bilayer system is obtained. It is worth mentioning that merons and anti-merons which combine in antiferromagnetic sublattices can also give rise to spin textures with fractional topological charge as predicted recently\cite{Gao2020}. For the AA stacking pattern, there are no additional triangular plaquettes for the entire bilayer system since the two layers are exactly on top of each other. Therefore, the topological charge always remains an integer.

\subsection{Phase diagram of NiBr$_2$ bilayer}
\begin{figure}
    \includegraphics[width=8.6 cm]{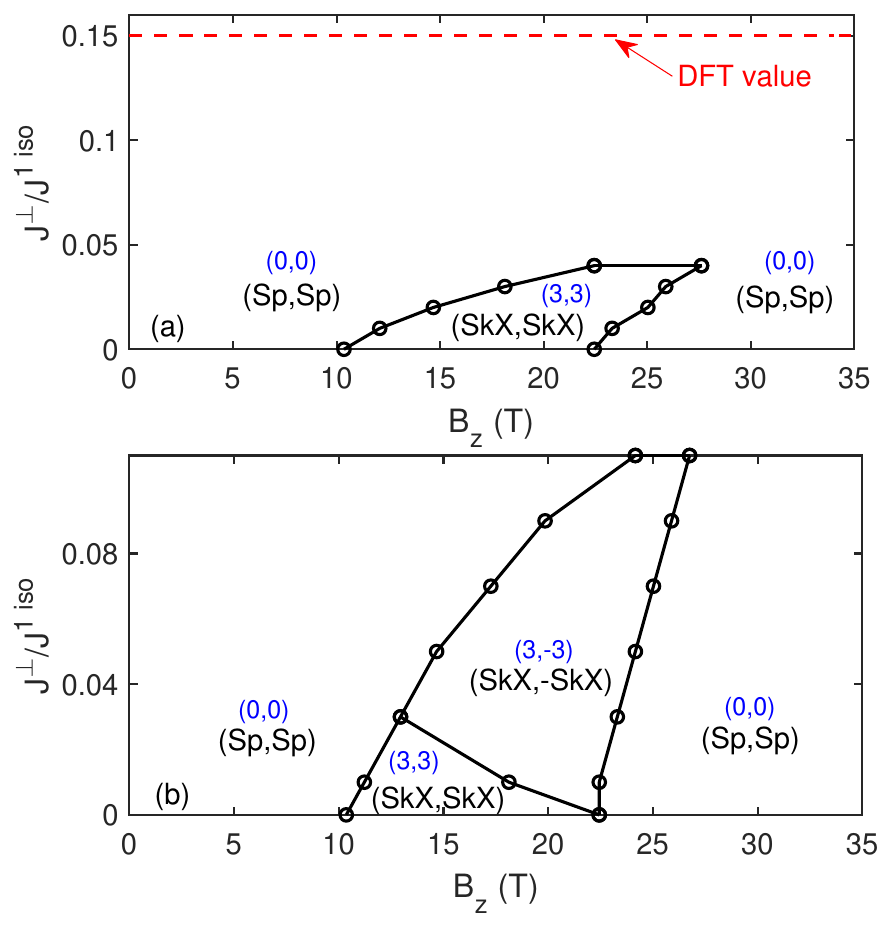}
    \caption{The phase diagram of NiBr$_2$ bilayer in (a) rhombohedral (b) AA stacking obtained from atomistic simulations. The dotted red line shows the DFT value. The area occupied by the SkX phase is very small compared to NiI$_2$ because of the absence of magnetic anisotropy. \cite{amoroso2020spontaneous}}
    \label{NiBr2_ab}
\end{figure}

Fig. \ref{NiBr2_ab}(a) shows the phase diagram of NiBr$_2$ in rhombohedral stacking. A small region of (SkX, SkX) phase coexists with (SkX, -SkX) phase with $\Delta E \sim 0.02$ meV. The skyrmion phase space is significantly smaller compared to NiI$_2$ due to the lack of anisotropic exchange in NiBr$_2$. \cite{amoroso2020spontaneous}. NiBr$_2$ does not exhibit antibiskyrmionic phases for the same reason. Similar to NiI$_2$, the antiferromagnetic coupling suppresses the SkX phase.  
The $J^{\perp}$ obtained from DFT \cite{pozzi2021singlelayer} implies that it may not be possible to observe skyrmions in bilayer NiBr$_2$ even in the presence of a magnetic field. The spin textures of the different phases are shown in the Appendix D.
Fig. \ref{NiBr2_ab} (b) shows the phase diagram of NiBr$_2$ in AA stacking. The 
(SkX, SkX) phase coexists with (SkX,-SkX) phase with $\Delta E \sim 0.02$ meV.

\subsection{Multiferrocity and Polarization}
NiI$_2$ undergoes a magnetic transition at $T_{c1} \sim 76$ K to an AFM state with FM planes. At $T_{c2} \sim 59.5$ K, a second magnetic transition to a helimagnetic (or Sp) phase takes place at which NiI$_2$ starts to exhibit a finite electric polarization\cite{KUINDERSMA1981231,BILLEREY1977138}. In the monolayer limit, the Sp phase and the polarization survivies but appears at a lower temperature  $T_{c2} \sim 21$ K. The origin of the polarization can be traced back to the non-colinear spin texture of the Sp phase\cite{Fumega_2022} and demonstrated via a Ginzburg-Landau approach. Under a time reversal symmetry operation, $t \rightarrow -t$, the polarization is unchanged, $\textrm{\textbf{P}}\rightarrow \textrm{\textbf{P}}$. Yet the magnetization flips, $\textrm{\textbf{M}} \rightarrow \textrm{\textbf{-M}}$. This requires the lowest order coupling between $\textrm{\textbf{P}}$ and $\textrm{\textbf{M}}$ to be quadratic in $\textrm{\textbf{M}}$. The symmetry with respect to the parity, $\textrm{\textbf{r}} \rightarrow \textrm{\textbf{-r}}$, transforms $\textrm{\textbf{P}} \rightarrow \textrm{\textbf{-P}}$ and $\textrm{\textbf{M}} \rightarrow \textrm{\textbf{M}}$ implies that a linear coupling in $\textrm{\textbf{P}}$ needs to contain one gradient of $\textrm{\textbf{M}}$. Therefore the lowest order coupling term between $\textrm{\textbf{P}}$ and $\textrm{\textbf{M}}$ has the form \cite{PhysRevLett.96.067601}
\begin{equation}
    \Phi_{em} (\textrm{\textbf{P}},\textrm{\textbf{M}}) = \gamma \textrm{\textbf{P}} \cdot [\textrm{\textbf{M}}(\nabla \cdot \textrm{\textbf{M}})-(\textrm{\textbf{M}} \cdot \nabla) \textrm{\textbf{M}}+ ...]
\end{equation}
The quadratic term in the electric part of the thermodynamic potential is $\Phi_{e}(\textrm{\textbf{P}}) =P^2/2 \chi_e$ where $\chi_e$ is the dielectric susceptibility. In order to determine $\textrm{\textbf{P}}$, we take the variation of $\Phi_e + \Phi_{em}$ with respect to $\textrm{\textbf{P}}$ which leads to 
\begin{equation}
    \textrm{\textbf{P}} = \gamma \chi_e  [\textrm{\textbf{M}}(\nabla \cdot \textrm{\textbf{M}})-(\textrm{\textbf{M}} \cdot \nabla) \textrm{\textbf{M}}]
    \label{p}
\end{equation}

\begin{figure}
    \includegraphics[width=8.6 cm]{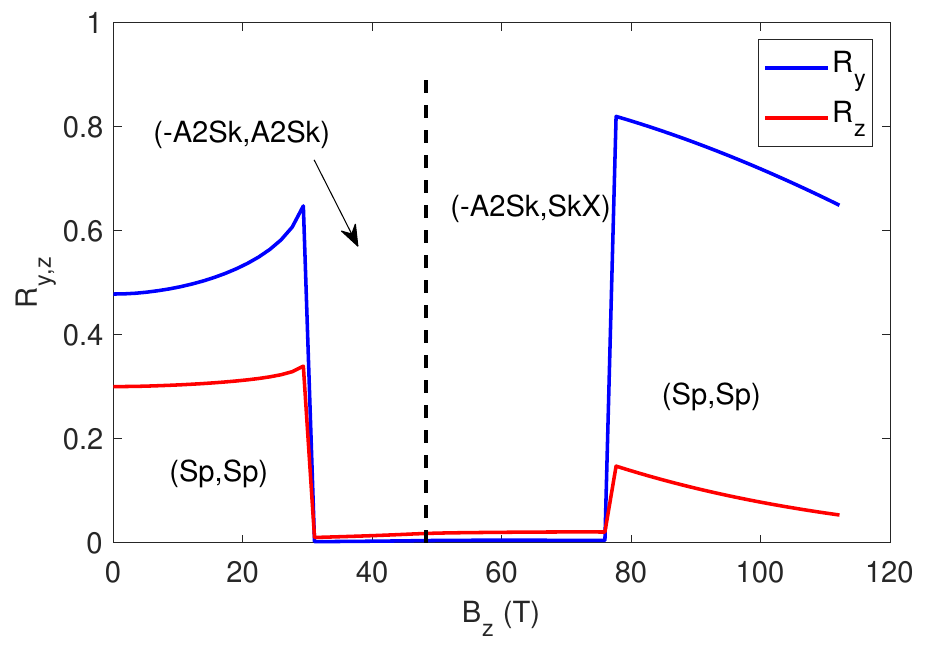}
    \caption{Induced polarization, R$_{y/z} = {\rm P}_{y/z}/\gamma \chi_e$ as a function of B for NiI$_2$ bilayer  ($J^{\perp}/J^{1\rm{iso}} = 0.3$). The polarization increases as with B in the spiral phase whereas it is negligible in the topologically nontrivial phases.}
    \label{pol1}
\end{figure}

We use eq.~\ref{p} to calculate $\textrm{\textbf{R}}=\textrm{\textbf{P}}/\gamma \chi_e$  for the spin textures obtained from LLG simulations. Since $\gamma$ and $\chi_e$ depend on material specific quantities, $\textrm{\textbf{R}}$ is not a good indicator for the magnitude of the induced polarization. However, it is effective to capture the trends as a function of tuning parameters such as B. For a coplanar spiral, it is straightforward to show that $\textrm{\textbf{P}} \propto (\hat{z}  \times \hat{\bf q})$ \cite{Fumega_2022, PhysRevLett.96.067601} where ${\bf q}$ is the ordering wave vector of the spiral. Therefore, coplanar spirals can only exhibit polarization that lies in the plane. In Fig. \ref{pol1}, we present the $\textrm{\textbf{R}}$ as a function of B for $J^\perp/J^{1\rm{iso}} =0.3$ in rhombohedral stacking. We find that in the Sp phase, both in-plane and $z$ component of $\textrm{\textbf{R}}$ is finite. This is due to the non-coplanar nature of the Sp phase. As a function of B, $\textrm{\textbf{R}}$ increases in the Sp phase. This effect has also been observed in experiments\cite{PhysRevB.87.014429}. On the contrary, the induced polarization in topologically non-trivial phases including SkX and A2Sk is negligible compared to the Sp phase. As direct observation of skyrmions in 2D systems is a challenging task, polarization can be used as an indirect probe to detect topologically non-trivial phases.

\section{Conclusions}

We studied the magnetic phases of bilayer NiI$_2$ and NiBr$_2$ in both rhombohedral and AA stacking via atomistic simulations. For both materials, we find that interlayer exchange strongly suppresses the SkX phase. In NiI$_2$, the depleted region is occupied by antibiskyrmionic phases with varying topological charges. We provide a detailed analysis for the competition between these phases. Due to weak exchange anisotropy in NiBr$_2$, interlayer exchange quickly destroys the topologically nontrivial phases and leads to the Sp phase. We conclude with an analysis on the induced polarization due to non-coplanar magnetic textures and show that the topological phases exhibit negligible polarization. Interesting future directions include moir\'e superlattices of helimagnets and skyrmions in magnetically frustrated systems.

\section{Acknowledgements}
This work is supported by NSF Award No. DMR 2206987. MA acknowledges support from Fulbright Scholarship.


\appendix
\begin{table}[!h]
\begin{center}
\begin{tabular}{||c | c |c | c |c |c |c |c |c  | c ||} 
 \hline
  & J$^{1 \rm{iso}}$ & J$^{2 \rm{iso}}$ & J$^{3 \rm{iso}}$ & J$_{x x}$ & J$_{y y}$ & J$_{z z}$ & J$_{yz}$ & J$_{x z}$ & J$_{x y}$\\ [0.5ex] 
 \hline\hline
 NiI$_2$ & -7.0 & -0.3 & 5.8 & -1.0 & 1.4 & -0.3 & -1.4 & 0 & 0\\ 
 \hline
 NiBr$_2$ & -5.9 & -0.1 & 2.9 & -0.1 & 0.1 & 0 & -0.1 & 0 & 0 \\ [1ex] 
 \hline
\end{tabular}
\caption{\label{table1} Intralayer exchange parameters for NiI$_2$ and NiBr$_2$\cite{amoroso2020spontaneous}.}
\end{center}
\end{table}

\section{Magnetic exchange parameters}
\begin{figure}
    \includegraphics[width=8.4 cm]{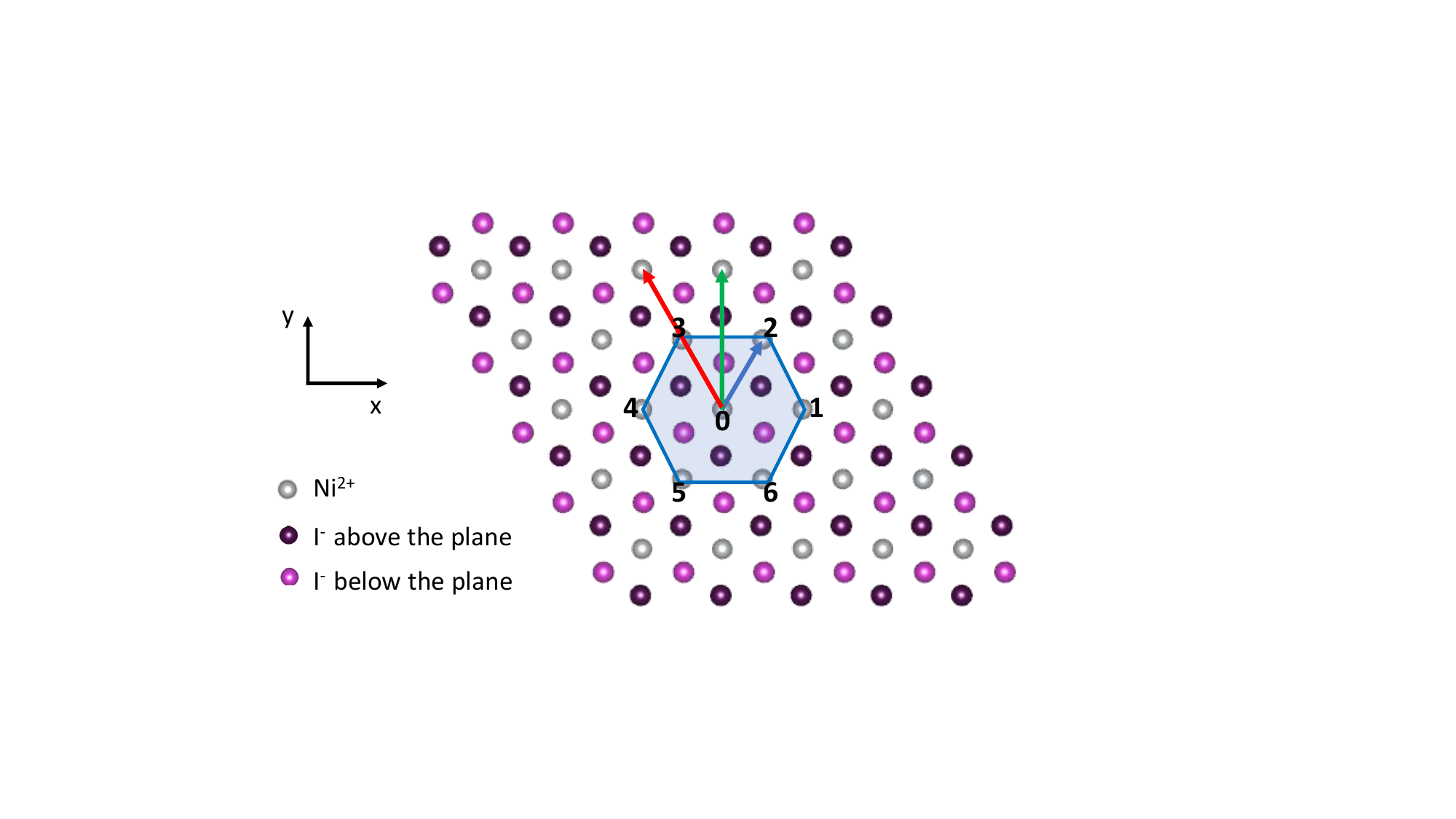}
    \caption{The nearest neighbours of monolayer NiI$_2$ are numbered in black and enclosed within the blue hexagon. The blue green and red arrows denote the 1st nearest, 2nd nearest and 3rd nearest neighbours respectively.}
    \label{lat3}
\end{figure}

For the intra-layer spin Hamiltonian, we use the magnetic exchange parameters that are obtained from first principle calculations by Ref. \citenum{amoroso2020spontaneous} shown in Table II. These couplings are for the Ni$_0$-Ni$_1$ pair (see Fig. \ref{lat3}) whose bonding vector is chosen parallel to the Cartesian $x$ axis and given by:

\begin{equation}
{\bf{J}}^{(0^{\circ})}=
\begin{pmatrix}
\rm{J}_{\rm{xx}} & 0 & 0 \\
0 & \rm{J}_{\rm{yy}} & \rm{J}_{\rm{yz}} \\
0 & \rm{J}_{\rm{yz}} & \rm{J}_{\rm{zz}} 
\end{pmatrix} \, ,
\label{eq:D1}
\end{equation}

\noindent where, by symmetry, $\rm{J}_{\rm{zy}}=\rm{J}_{\rm{yz}}$ and the other off-diagonal terms are nominally zero. The corresponding tensor for the symmetry-equivalent pairs Ni$_0$-Ni$_3$ and Ni$_0$-Ni$_5$ rotated by $\pm$120$^{\circ}$
can be deduced by exploiting the three-fold rotational symmetry, leading to: 

\begin{equation}
{\bf{J}}^{(\frac{2\pi}{3})}=
  \begin{pmatrix}
  \frac{1}{4}(\rm{J}_{\rm{xx}} + 3\rm{J}_{\rm{yy}}) & - \frac{\sqrt{3}}{4}(\rm{J}_{\rm{xx}} - \rm{J}_{\rm{yy}}) & -\frac{\sqrt{3}}{2}\rm{J}_{\rm{yz}} \\ \\
-\frac{\sqrt{3}}{4}(\rm{J}_{\rm{xx}} - \rm{J}_{\rm{yy}}) & \frac{1}{4}(3J_{xx} + \rm{J}_{\rm{yy}}) & -\frac{1}{2}\rm{J}_{\rm{yz}} \\ \\
-\frac{\sqrt{3}}{2}\rm{J}_{\rm{yz}}  & -\frac{1}{2}\rm{J}_{\rm{yz}}  & \rm{J}_{\rm{zz}}
  \end{pmatrix} \, .
\label{eq:D2}
\end{equation}

\section{Skyrmions and Topological charge}\label{appa}
\begin{figure}
    \includegraphics[width=8.6 cm]{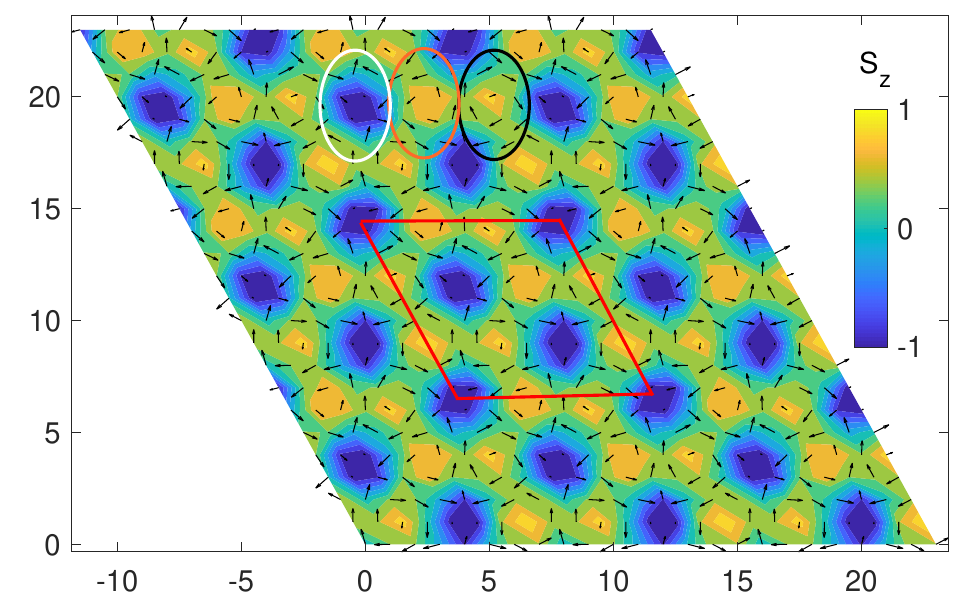}
    \caption{The -A2Sk phase of NiI$_2$ bilayer in rhombohedral stacking at B$_z$ = 31.1 T and J$^{\perp}$ = 0.1 J$^{1iso}$. The area marked by red denotes the magnetic unit cell. The white ellipse denotes the antibiskyrmion vortex of type V$_2$ while the orange and black ellipses denote the anticlockwise skyrmion and clockwise skyrmion vortices of type V$_1$ respectively.}
    \label{A2sk}
\end{figure}

We use the following definition \cite{nagaosa2013topological} for the topological charge of a continuous field $\mathbf{s}(x,y)$

\begin{equation}
    Q=\frac{1}{4 \pi} \int d^2 \mathbf{r} \; \mathbf{s} \cdot \bigg( \frac{\partial \mathbf{s}}{\partial x} \times \frac{\partial \mathbf{s}}{\partial y} \bigg)
    \label{tQ_main}
\end{equation}

Physically, this signifies the number of times the spins wrap around a unit sphere. Substituting $\mathbf{s}=(\cos \Phi(\phi)\, \sin \Theta(r), \sin \Phi(\phi) \sin \Theta(r) , \cos \Theta(r))$ and  $\mathbf{r}=(r \cos \phi, r \sin \phi)$ in eq. \ref{tQ_main} we get
$Q= -\frac{1}{4 \pi}
  [\cos \Theta(r)] \Bigr|_{\theta (r=0)}^{\theta (r=R)} \, 
  [\Phi] \Bigr|_{\phi=0}^{\phi=2 \pi}$ . Therefore $Q$ is the product of polarity which is the first part and vorticity given by $\omega= [\Phi] \Bigr|_{\phi=0}^{\phi=2 \pi} / 2 \pi$ \cite{Amoroso_Nanomatt2021} which is the second part.
  $R$ is defined as the radius of a skyrmion. The in-plane component of the spins wrap around twice in biskyrmions leading to twice the vorticity (the $\phi$ term) compared to skyrmions. The topological charge is evaluated using the definition \cite{BERG1981412} of a discrete lattice model where we sum over $\Omega$ defined as 

  \begin{equation}
    \rm \tan \left( \frac{\Omega}{2} \right)=\frac{\mathbf{s}_{1} \cdot \mathbf{s}_{2} \times \mathbf{s}_{3}}{1+\mathbf{s}_{1} \cdot \mathbf{s}_{2}+\mathbf{s}_{2} \cdot \mathbf{s}_{3}+\mathbf{s}_{3} \cdot \mathbf{s}_{1}}
    \label{appb1}
\end{equation}
over a magnetic unit cell. Here $\Omega$ is calculated over a triangular plaquette with spins $\mathbf{s}_{1}$, $\mathbf{s}_{2}$ and $\mathbf{s}_{3}$.

\begin{figure}[h!]
    \includegraphics[width=8 cm]{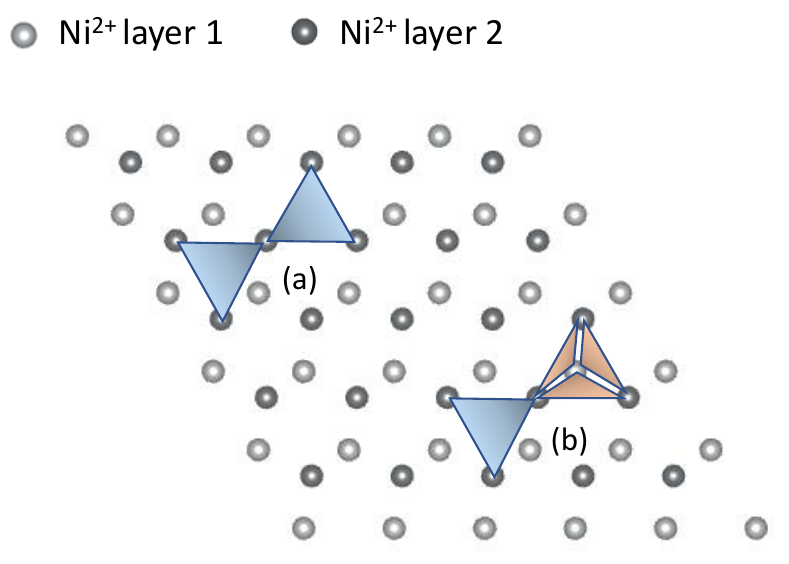}
    \caption{Two ways of calculating the topological charge. (a) Triangular plaquettes for individual layers. (b) Triangular plaquettes for the whole bilayer system}
    \label{lat2}
\end{figure}

A representative spin texture of -A2Sk phase is shown in Fig. \ref{A2sk}. The antibiskyrmionic vortex of type V$_2$ as defined in the main text (see subsection III B1) is shown by a white ellipse in Fig. \ref{A2sk} whereas the two skyrmionic vortices of type V$_1$ are denoted by a black and an orange ellipse. These form a unit which give an integer topological charge $Q=-2$ when evaluated using the above formula for $Q$. Three of these units form a magnetic unit cell (area under the red parallelogram) with a total topological charge $Q=-6$\cite{amoroso2020spontaneous}.

We also use the spin structure factor to determine the magnetic phases in addition to the topological charge. It is defined as
\begin{equation}
S(\mathbf{q})=\frac{1}{N} \sum_{\alpha = x,y,z} \left\langle \left| \sum_{\mathrm{i}} s_{\mathrm{i},\alpha} e^{-i \mathbf{q} \cdot \mathbf{r}_{\mathrm{i}}} \right|^{2} \right\rangle ,
\end{equation}

where $N=L^{2}$ is the total number of spins and where the position of spin $s_{\mathrm{i}}$ is denoted by $\mathbf{r}_{\mathrm{i}}$.

\section{LLG simulations and the Luttinger-Tisza method}

The ground state of Hamiltonian (eq.~\ref{eq:1}) is determined by solving the Landau-Lifshitz-Gilbert (LLG) equation: \cite{1353448}
\begin{eqnarray} \label{eq:E1}
\frac{d\textbf{s}}{dt}=-\gamma \textbf{s}\times \textbf{B}^{\rm eff}+\alpha \textbf{s} \times \frac{d\textbf{s}}{dt} \, ,
\end{eqnarray}
where $\textbf{B}^{\rm eff}=-\delta H/\delta \textbf{s}$, $\gamma$ is the gyromagnetic ratio and $\alpha$ is Gilbert damping coefficient. We have solved the LLG equations self-consistently by keeping  $|{\bf{s}}|=1$ and imposing periodic boundary conditions. A semi-implicit midpoint algorithm \cite{Mentink_2010} was used in order to implement these equations because of its relative simplicity and the fact that the spins ${\bf{s}}$ do not have to be normalized after each step. For a particular magnetic field and interlayer coupling the lowest energy spin configuration was selected after converging around 200 simulations with random initial spin configurations.

The Luttinger-Tisza method \cite{Luttinger_PR1946} was used in order to get an estimate of the size of the magetic unit cell $(L \times L)$. This method replaces the hard spin constraint $|{\bf{s}}_i| = 1$ with a soft spin constraint $\sum_i |\mathbf{s}_{i}|^2 = N$ and enables us to determine the lowest energy, coplanar spiral configurations with a wave vector $\bf{q}$. This sets a natural length scale in the problem $L\sim 2\pi/q$ which is also important for determining the SkX and A2Sk phases as they are a superposition of three spirals with the same $q$ but rotated by 120 degrees with respect to each other. 

For an isotropic model with only the first and third nearest neighbour interactions $\mathrm{J^{1iso}}$ and $\mathrm{J^{3iso}}$, an analytical expression for the wave vector can be obtained: $q=2 \cos^{-1}[(1+\sqrt{1-2 \mathrm{J^{1iso}}/\mathrm{J^{3iso}}})/4]$ \cite{PhysRevB.93.184413,Batista_2016}. In LLG simulations, we consider a multiple of integer $L$. We  benchmarked the validity of this method via considering different system sizes in the simulations. We deduce that the Luttinger-Tisza method provides the correct $L$ in all cases.  

\section{The semi-implicit midpoint method}
The Landau-Lifshitz-Gilbert equation \ref{eq:E1} can be simplified to the Landau-Lifshitz form which is written as 
\begin{equation} \label{eq:E2}
\frac{d\textbf{s}}{dt}=\gamma_L \: \textbf{B}^{\rm eff} \times \textbf{s} +\gamma_L \alpha \; (\textbf{s}\times \textbf{B}^{\rm eff}) \times \textbf{s} 
\end{equation}
Here $\gamma_L = \frac{\gamma}{1 + \alpha^2}$ is the renormalized gyromagnetic ratio.
We can further simplify it to the form
\begin{equation}
    \frac{\partial \textbf{s}_i}{\partial t} = \textbf{a}_i(t,\{ \textbf{s}_j(t)\}) \times \textbf{s}_i (t)
\end{equation}
Here $\textbf{a}_i$ contains $\textbf{B}_i^{\rm eff}$ and other constants which can be taken to be 1. And $i$ denotes each site. This equation is solved using a predictor-corrector step. The predictor step is given as 
\begin{equation}
    \textbf{s}^p_i(t+\delta t) = \textbf{s}_i(t) + \textbf{a}_i(t,\{ \textbf{s}_j(t)\}) \times \delta t \: \frac{\textbf{s}_i(t)+\textbf{s}^p_i(t+\delta t)}{2}
\end{equation}
Rearranging this gives
\begin{equation}
    \textbf{s}^p_i(t+\delta t)-\frac{\delta t}{2} \textbf{a}_i  \times \textbf{s}^p_i(t+\delta t) = \textbf{s}_i(t) + \frac{\delta t}{2} \: \textbf{a}_i \times \textbf{s}_i(t)
\end{equation}
For each site $i$ the above equation can be written in matrix form as 
\begin{equation}
    \textbf{A S}= \textbf{B}
\end{equation}
where $\textbf{A} = \textbf{I} - \begin{pmatrix}
0 & -a_z & a_y\\
a_z & 0 & -a_x \\
-a_y & a_x & 0 
\end{pmatrix}$, $\textbf{S}= \begin{pmatrix}
 s_x^p\\
 s_y^p\\
s_z^p
\end{pmatrix}$ and $\textbf{B}= \textbf{s}_i(t) + \frac{\delta t}{2} \textbf{a}_i \times \textbf{s}_i(t)$. Since this is a linear equation, it can be easily solved using   `linsolve' command in MATLAB. The `parfor' command can be used to parallelly compute the spins at different sites and speed up the convergence. The calculated spins $\textbf{s}^p_i$ are now used to generate new values of $\textbf{a}_i$. This is used in the corrector step which is given by
\begin{align}
    \textbf{s}_i(t+\delta t) =& \textbf{s}_i(t) + \textbf{a}_i \Big( t+\frac{\delta t}{2},\Big\{ \frac{\textbf{s}_j(t)+\textbf{s}_j^p(t+\delta t)}{2} \Big\} \Big) \nonumber \\ 
    &\times \delta t \: \frac{\textbf{s}_i(t)+\textbf{s}^p_i(t+\delta t)}{2}
\end{align}
The corrector step is also a linear equation and is solved the same way as the predictor step. The code converges when the difference between the spins at consecutive time steps is $ \leq 10^{-5}$. 

\section{Spin textures of NiBr$_2$ bilayer}
\begin{figure}[h!]
    \includegraphics[width=8.6 cm]{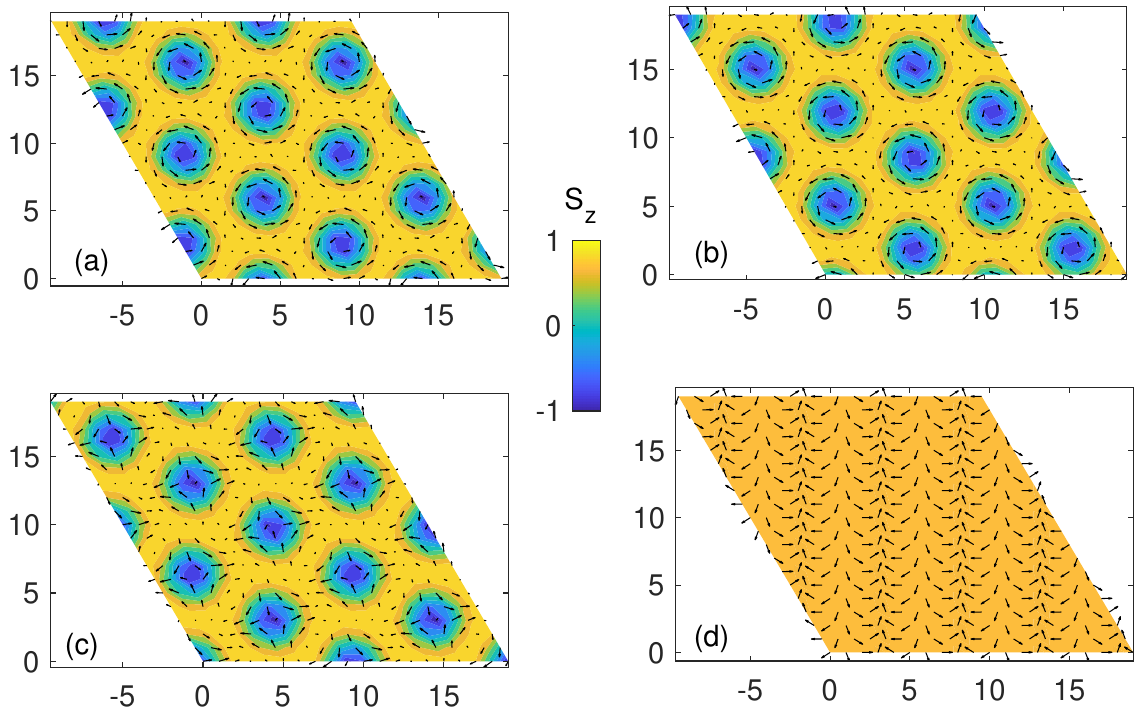}
    \caption{Spin textures for NiBr$_2$ bilayer for J$_{\perp}/J^{1\rm{iso}} = 0.01$ and B$_z$= 17.2 T. (a) and (b) show the skyrmionic phases of the two layers. Notice that the skyrmions are not on top of each other but are adjusted in a way so as to minimize the total energy. (c) shows antiskyrmionic phase and (d) shows that spiral phase. For the parameters considered, (a) and (b) are the ground state and (c) and (d) are local minima.} 
    \label{br2_sk}
\end{figure}

In Fig.~\ref{br2_sk}, we present the different spin textures of bilayer NiBr$_2$ obtained from the LLG simulations.

%

\end{document}